\documentclass{PoS}

\leftline{\parbox{3cm}{\large\rm DESY~09-034 ~~LPT-Orsay~09-32}
\vspace{1cm}}

\title{Partially quenched study of strange baryon with Nf = 2 twisted mass fermions }

\ShortTitle{Partially quenched study of strange baryon with Nf = 2 twisted mass fermions }

\author{\speaker{Vincent Drach}\thanks{On behalf of the European
       Twisted Mass Collaboration}, Mariane Brinet, Jaume Carbonell,
       Zhaofeng Liu    \\  
     Laboratoire de Physique Subatomique et de Cosmologie,
       UJF, CNRS/IN2P3, INPG \\
        E-mail: \email{drach@lpsc.in2p3.fr, mariane@lpsc.in2p3.fr,
       carbonel@lpsc.in2p3.fr, liu@lpsc.in2p3.fr}} 
\author{Remi Baron, Pierre Guichon\\
 CEA, Centre de Saclay, IRFU/Service de Physique Nucléaire, F-91191 
Gif-sur-Yvette, France
E-mail: \email{remi.baron@cea.fr, pierre.guichon@cea.fr}}
\author{Olivier P\`ene\\
Laboratoire de Physique Th\'eorique,
               UMR8627, Universit\'e Paris XI, 91405 Orsay-Cedex, France\\
Email: \email{Olivier.Pene@th.u-psud.fr}}
\author{Constantia Alexandrou, Tomasz Korzec, Giannis Koutsou\\
  Department of Physics, University of Cyprus, P.O. Box 20537,
 1678 Nicosia, Cyprus\\
        E-mail: \email{alexand@ucy.ac.cy, korzec@ucy.ac.cy, koutsou@ucy.ac.cy}}

\author{Elisabetta Pallante, Siebren Reker\\
Centre for Theoretical Physics, University of Groningen, 974
7 AG, Netherlands \\ 
 E-mail:: \email{e.pallante@rug.nl, s.f.reker@rug.nl }}
\author{Carsten Urbach\\
Institut f\"ur Elementarteilchenphysik, Fachbereich Physik, Humbolt Universit\"at zu Berlin, D-12489, Berlin, Germany\\
Email: \email{Carsten.Urbach@physik.hu-berlin.de}}
 \author{K.~Jansen\\
NIC, DESY, Zeuthen, Platanenallee 6, D-15738 Zeuthen, Germany\\
Email \email{Karl.Jansen@desy.de}}

\abstract{ 

We present results on the mass of the  baryon octet and 
decuplet using two flavors of light dynamical twisted mass fermions.
The strange quark mass is fixed to its physical value from the kaon
sector in a partially quenched set up. 
Calculations are performed for light quark masses corresponding to a pion 
mass in the range  270-500 MeV  and lattice sizes of 2.1~fm and 2.7~fm. 
We check for cut-off effects and isospin breaking by evaluating the baryon masses at two
different lattice spacings.
We carry out  a chiral extrapolation for the octet baryons and 
discuss results for the $\Omega$.

}

\FullConference{The XXVI International Symposium on Lattice Field Theory \\
                 July 14 - 19, 2008\\
                 Williamsburg, Virginia, USA}

\begin{document}

\newcommand{\be}{\begin{equation}}
\newcommand{\ee}{\end{equation}}
\newcommand{\beq}{\begin{eqnarray}}
\newcommand{\eeq}{\end{eqnarray}}

\section{Introduction} 
Twisted mass fermions provide a promising formulation of lattice QCD that
allows for automatic ${\cal O}(a)$ improvement, infrared regularization
of small
eigenvalues and fast dynamical simulations~\cite{TM intro}. This work
 is an extension of the study on the  nucleon
and $\Delta$ masses~\cite{nucleon} to the strange baryon sector. It uses
two degenerate dynamical twisted mass fermions ($N_F=2$) 
and a strange quark in the partially quenched approximation. 
The octet and decuplet baryon masses are computed at several pion masses.

We use the tree-level Symanzik improved gauge action and work
at maximal twist to realize ${\cal O}(a)$-improvement.
The fermionic action for two degenerate flavors of quarks
 in twisted mass QCD is given by
\be
S_F= a^4\sum_x  \bar{\psi}(x)\bigl(D_W[U] + m_0
+ i \mu \gamma_5\tau^3  \bigr ) \psi(x)
\label{S_tm}
\ee
with $D_W[U]$ the massless Wilson Dirac operator. The parameter $m_0$
is adjusted such that $\mu$ represents the bare quark mass~\cite{Boucaud:2008xu} .
The twisted mass term in the fermion action of Eq.~(\ref{S_tm})
 breaks isospin symmetry since the u- and d- quarks differ
 by having opposite signs
for the  $\mu$-term. This isospin breaking is a
cutoff effect of ${\cal O}(a^2)$. 

In all the results presented here, the lattice spacing $a$ 
as well as
the strange quark mass has been fixed in the meson sector ($f_{\pi}
$~\cite{Urbach} and $m_K$~\cite{fkaon}). Our work will confirm the consistency between
meson and baryon description in the partially quenched approximation.
It is important to note that no new parameter has been tuned in this study.

\section{Lattice techniques}

The input parameters of the calculation (L, $\beta$ and $\mu$) are
collected in Table~\ref{Table:params}. The corresponding lattice
spacing $a$ and the pion mass values are taken from ~\cite{Urbach} .
They span a pion mass range from 270 to 500~MeV. At $m_{\pi}\approx 300$ MeV
we have simulations for lattices of
 spatial size, $L_s=2.1$~fm and $L_s=2.7$~fm at $\beta=3.9$ allowing to
investigate finite size effects. We provide a preliminary check of 
finite lattice spacing effects by comparing results at $\beta=3.9$ and $\beta=4.05$. 
The masses of the octet and decuplet are extracted from two-point
correlators using the standard interpolating fields (see e.g ~\cite{Leinweber}) and errors are estimated with the jackknife method.\\

Local interpolating fields  are not optimal for suppressing excited state contributions. We instead apply
 Gaussian smearing to each  quark field,
 $q({\bf x},t)$:
$q^{\rm smear}({\bf x},t) = \sum_{\bf y} F({\bf x},{\bf y};U(t)) q({\bf y},t)$
using the gauge invariant smearing function  
\be 
F({\bf x},{\bf y};U(t)) = (1+\alpha H)^ n({\bf x},{\bf y};U(t)),
\ee
constructed from the
hopping matrix,
$ 
H({\bf x},{\bf y};U(t))= \sum_{i=1}^3 \biggl( U_i({\bf x},t)\delta_{{\bf x,y}-i} +  U_i^\dagger({\bf x}-i,t)\delta_{{\bf x,y}+i}\biggr).
$
 Furthermore we apply APE smearing to the spatial links  that enter the hopping matrix.
The parameters for the Gaussian and APE smearing are the same as those  used in our previous work devoted to the nucleon~\cite{nucleon}.

For a partially quenched strange quark  we use a  Osterwalder-Seiller  fermion, defined by the  action :
\be
S_{OS} = a^4\sum_x  \bar{s}(x)\bigl(D_W[U] + i \mu_s \gamma_5  \bigr ) s(x)
\label{S_os}
\ee
The bare mass $ \mu_s$ is tuned using the mass of the kaon at the
physical point~\cite{fkaon}. 

\begin{table}[h]
\begin{center}
\begin{tabular}{c|cccccc}
\hline
\multicolumn{6}{c}{$\beta=3.9$, 
$a=0.0855(6)$~fm from $f_\pi$~\cite{Urbach}}\\
\hline 
 $24^3\times 48$, $L_s=2.1$~fm &$a\mu$         & 0.0030   &    0.0040      &   0.0064     &  0.0085     &   0.010 \\ 
 & Stat. & -  &795 &547 & 348 &477 \\ 
&$m_\pi$~(GeV) & -  &0.3131(16) & 0.3903(9) & 0.4470(12) & 0.4839(12)\\

$32^3\times 64$, $L_s=2.7$~fm &$a\mu$ & 0.003 & 0.004 & & & \\
& Stat. & 133  &101 & & & \\
&  $m_\pi$~(GeV)& 0.2696(9)   & 0.3082(6) & & &  \\

\\\hline
\hline
\multicolumn{6}{c}{ $\beta=4.05$, $a=0.0666(6)$~fm from $f_\pi$~\cite{Urbach}}\\
\hline
$32^3\times 64$, $L_s=2.1$~fm &$a\mu$         & 0.0030     & 0.0060     & 0.0080     & 0.012\\
& Stat.   &138 &126 &  113 &182\\
 &$m_\pi$~(GeV) & 0.3070(18) & 0.4236(18) & 0.4884(15) & 0.6881(18) \\

\hline
\end{tabular}
\caption{The parameters of our calculation.}
\label{Table:params}
\end{center}
\vspace*{-0.8cm}
\end{table}

\section{Octet baryon masses}
\begin{figure}[ht!]   
\begin{minipage}{7.5cm}
\ifpdf
{\mbox{\includegraphics[height=5.5cm,width=7.5cm]{./plots/meff/meff_lambda_LL_SS}}}
\else
{\mbox{\includegraphics[height=5.5cm,width=7.5cm]{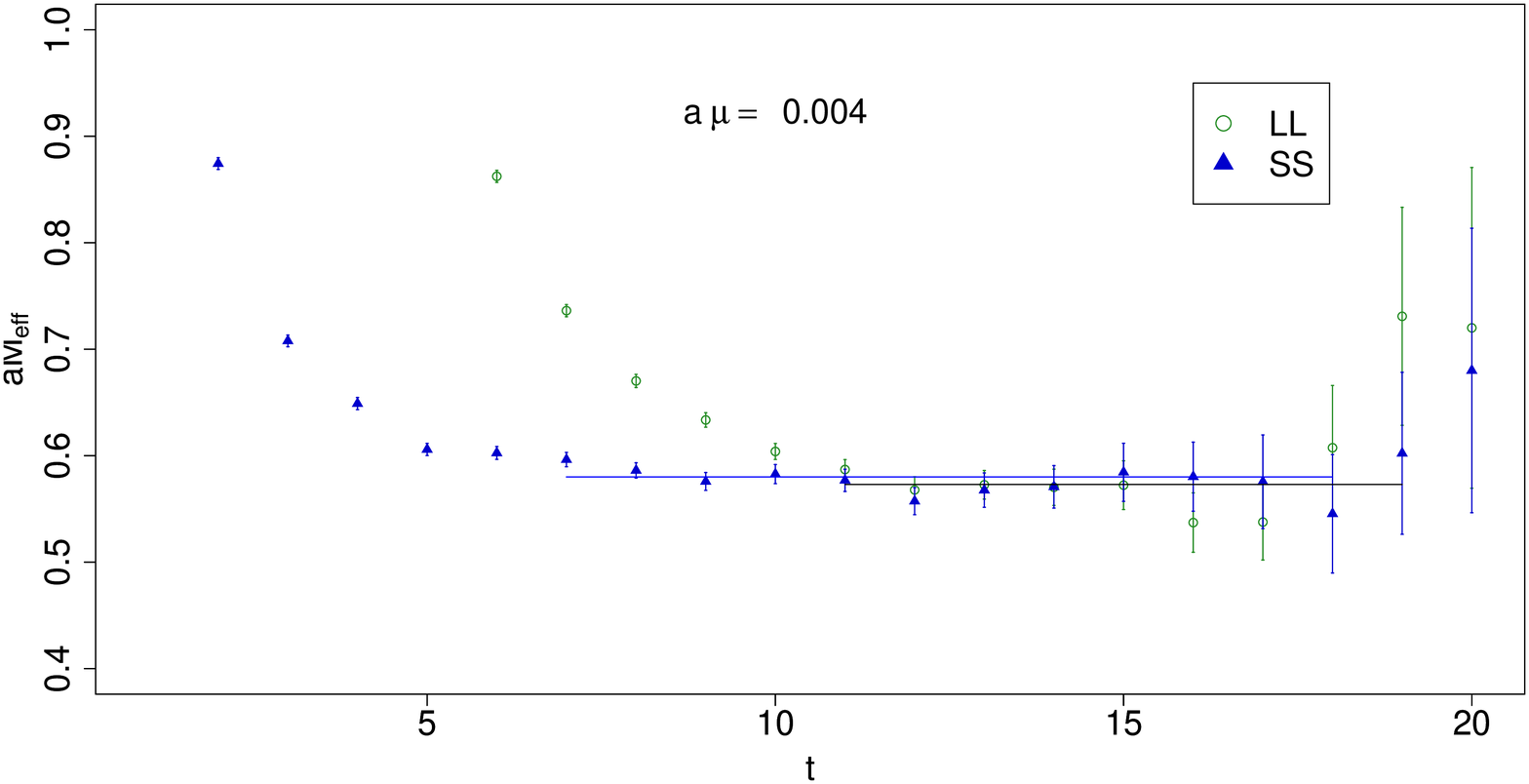}}}
\fi
\end{minipage}\hspace*{0.5cm}
\begin{minipage}[h]{7cm}
As can be seen, the excited state contributions  in the  SS correlators are suppressed,
yielding a constant asymptote in the effective mass a couple of time slices earlier.
 This allows  a better identification of the plateau 
and leads to a better accuracy in the mass extraction.
 The same kind of improvement is observed for all the octet
 and the decuplet states.
\end{minipage}
\caption{Plateaus for $\Lambda$  for $\beta=3.9$ ($a \approx
  0.085~\textit{fm}$) on a $24^3\times 48$ lattice for local-local (LL) and smeared-smeared (SS) correlators at a pion mass 
of $313~\textit{MeV}$.} 
\label{fig:lambdaLL_SS}
\end{figure}

We illustrate the quality of the plateaus that we  obtain   for the case of $\Lambda$ in
 Fig.~\ref{fig:lambdaLL_SS}, where we compare the effective masses 
computed using local source-local sink (LL) and  smeared source-smeared sink (SS)  correlators. 


\subsection{Chiral extrapolation}
As we will demonstrate in the next section, 
lattice artifacts are small. This  allows to perform the chiral 
extrapolation of  our data to the physical point at fixed lattice spacing. 
For the current discussion we do not distinguish between the
different isospin components of $\Sigma$ and $\Xi$ and present results averaging over the corresponding correlators. 
Unless we mention otherwise we use the
notation $\Sigma$ and $\Xi$ to denote the average of  the $(\Sigma^{+},\Sigma^0,\Sigma^-)$ and $(\Xi^0,\Xi^-)$ multiplet mass.
We will discuss  isospin breaking in a separate section. 

\begin{figure}[h]
\begin{minipage}[ht]{7.5cm}
\ifpdf
{\mbox{\includegraphics[height=6cm,width=7.5cm]{./plots/xfit_octet_nuc_substraction}}}
\else
{\mbox{\includegraphics[height=6cm,width=7.5cm]{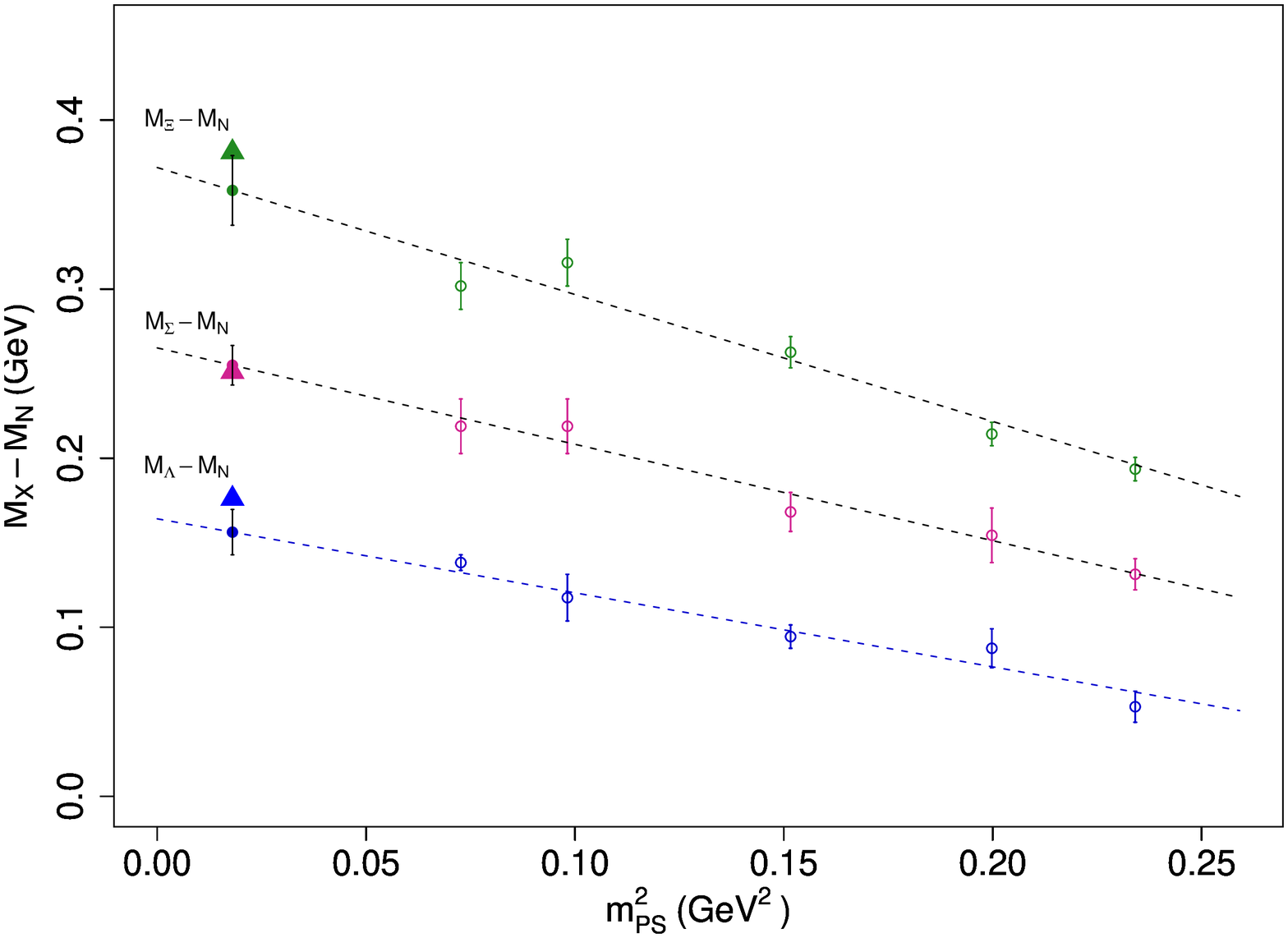}}}
\fi
\caption{Mass difference between members of the octet and the Nucleon
  extracted from ratio of correlators as a function of  $m_{\pi}^2$,
 at fixed lattice spacing ($\beta=3.9$, $a = 0.085~\textit{fm}$). Physical points are represented by triangles. }
\label{fig:nuc_sub_octet}
\end{minipage}
\hspace{0.5cm}
\begin{minipage}[ht]{7.5cm}\vspace*{-0.1cm}
\ifpdf
{\mbox{\includegraphics[height=6cm,width=7.5cm]{./plots/xfit_octet}}}
\else
{\mbox{\includegraphics[height=6cm,width=7.5cm]{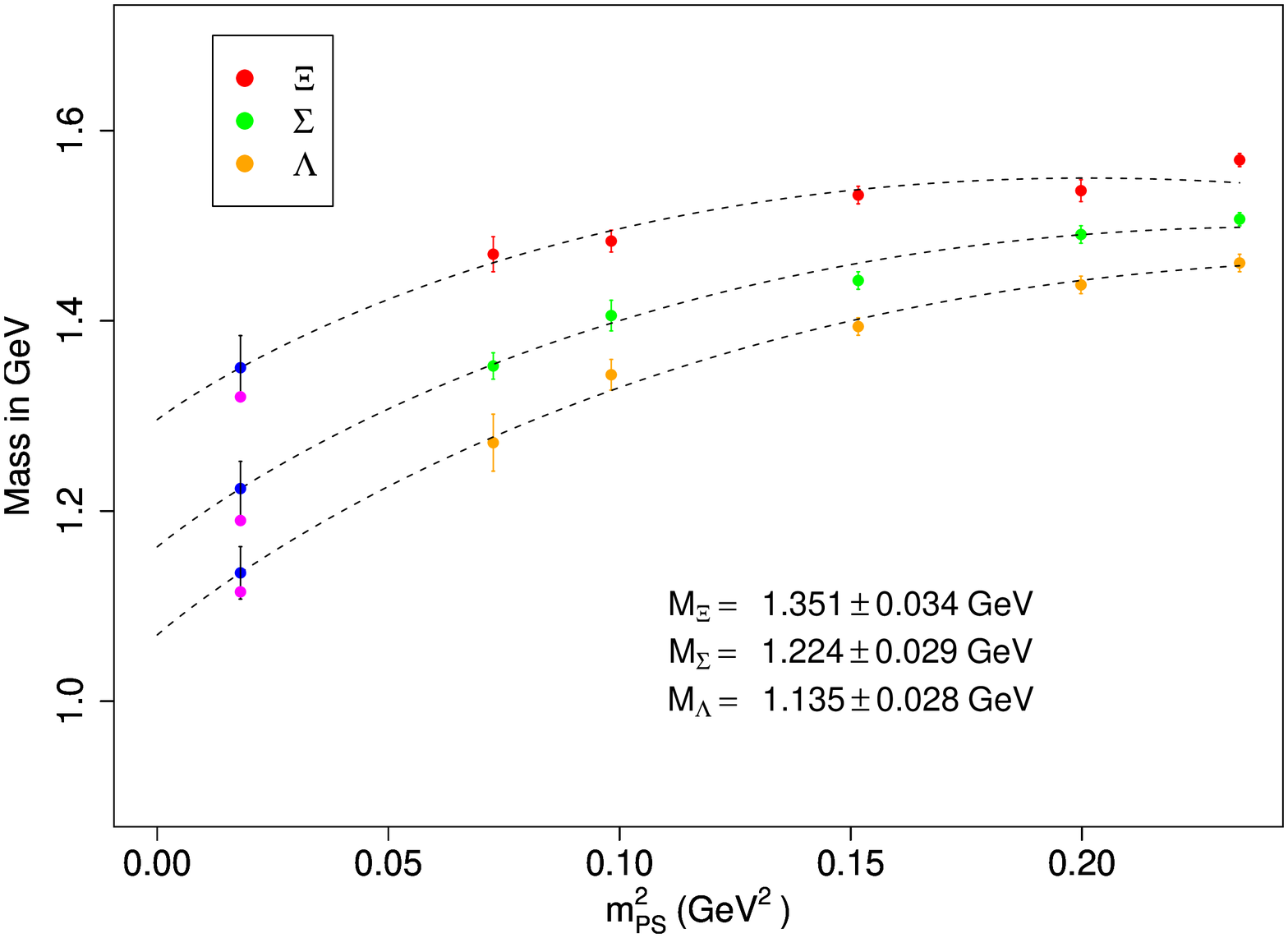}}}
\fi
\caption{Chiral fits of the  $\Lambda,\Sigma$ and $\Xi$ baryon masses. 
The experimental values for the masses  (shown in magenta) are not included in the fits. The masses obtained from extrapolating lattice data keeping
the cubic term fixed  are shown in blue with fixed.}
\label{fig:xfit_octet}
\end{minipage}
\end{figure}

We take the chiral expansion  of the baryon mass ($M_X$) in a partially 
quenched setup to leading order to be~\cite{qchipt} :

\be
M_{X} = M_0 +a_X m_{\pi}^2 + b_X m_{\pi}^3
\label{chiral}
\ee
Performing  a three-parameter  fit of our data is not stable. The  method
followed in the present analysis relies on the  observation displayed
in Fig.~\ref{fig:nuc_sub_octet}  that the
mass difference between any member of the octet and the nucleon is, within errors,
linear in $m_{\pi}^2$.
This suggests that the coefficient of the cubic term in Eq.~(\ref{chiral})
for the $\Lambda,\Sigma$ and $\Xi$ baryons is compatible with the nucleon one~\cite{chipt}, i.e :
\be
b_{N}= -\frac{3g_A^2}{32\pi f_{\pi}^2}.
\ee
Note that the SU(6) quark model, as well as the  SU(3) phenomenological analysis  of  the semi-leptonic decays and hyperon-nucleon interaction~\cite{Nagels,Nagels:1978} predict different values for this coefficient. 
In  Fig.~\ref{fig:xfit_octet}, we fix the cubic term to the nucleon one and fit the parameters $M_0$ and $a_X$ for the other states of the octet.
The mass values we find extrapolationa at the physical point are in good agreement  with the experimental results. 
A careful analysis of the cubic term contribution, including systematics, is however  required and will be detailed in a coming paper~\cite{drach}.

\clearpage
\subsection{Lattice artifacts and isospin breaking}
 
\begin{center} 
\begin{figure}[ht]
\begin{minipage}{7.5cm}
\ifpdf 
{\mbox{\includegraphics[height=6cm,width=7.5cm]{./plots/artefact.sigma}}}
\else
{\mbox{\includegraphics[height=6cm,width=7.5cm]{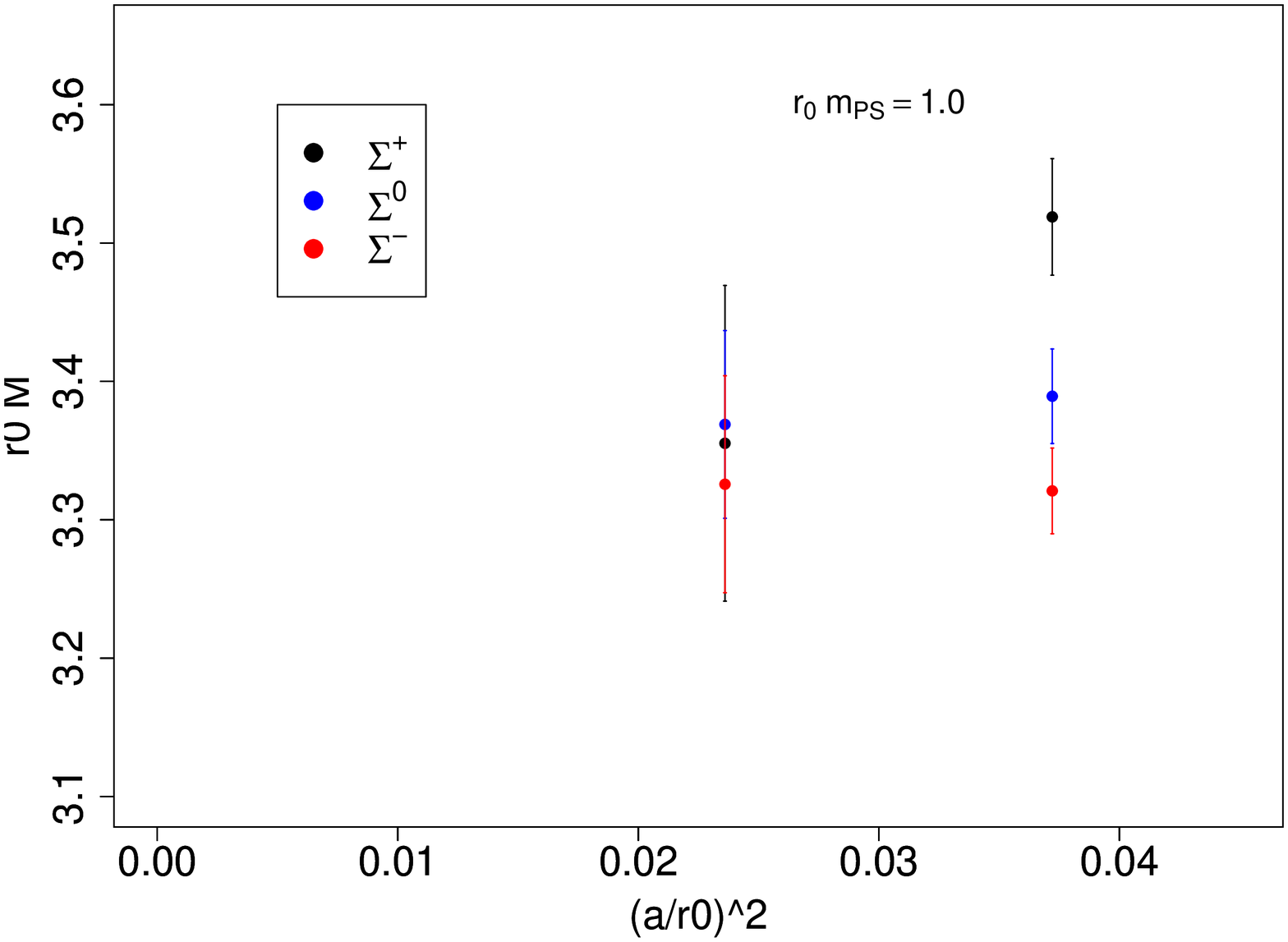}}}
\fi
\caption{$r_0 M_{\Sigma}$ as a function of $(a/r_0)^2$ for $\Sigma^+$, $\Sigma^0$ and  $\Sigma^-$.  }
\label{fig:artefact_sigma}
\end{minipage}\hspace*{0.3cm}
\begin{minipage}{7.5cm}
\ifpdf
{\mbox{\includegraphics[height=6cm,width=7.5cm]{./plots/artefact.xi}}}
\else
{\mbox{\includegraphics[height=6cm,width=7.5cm]{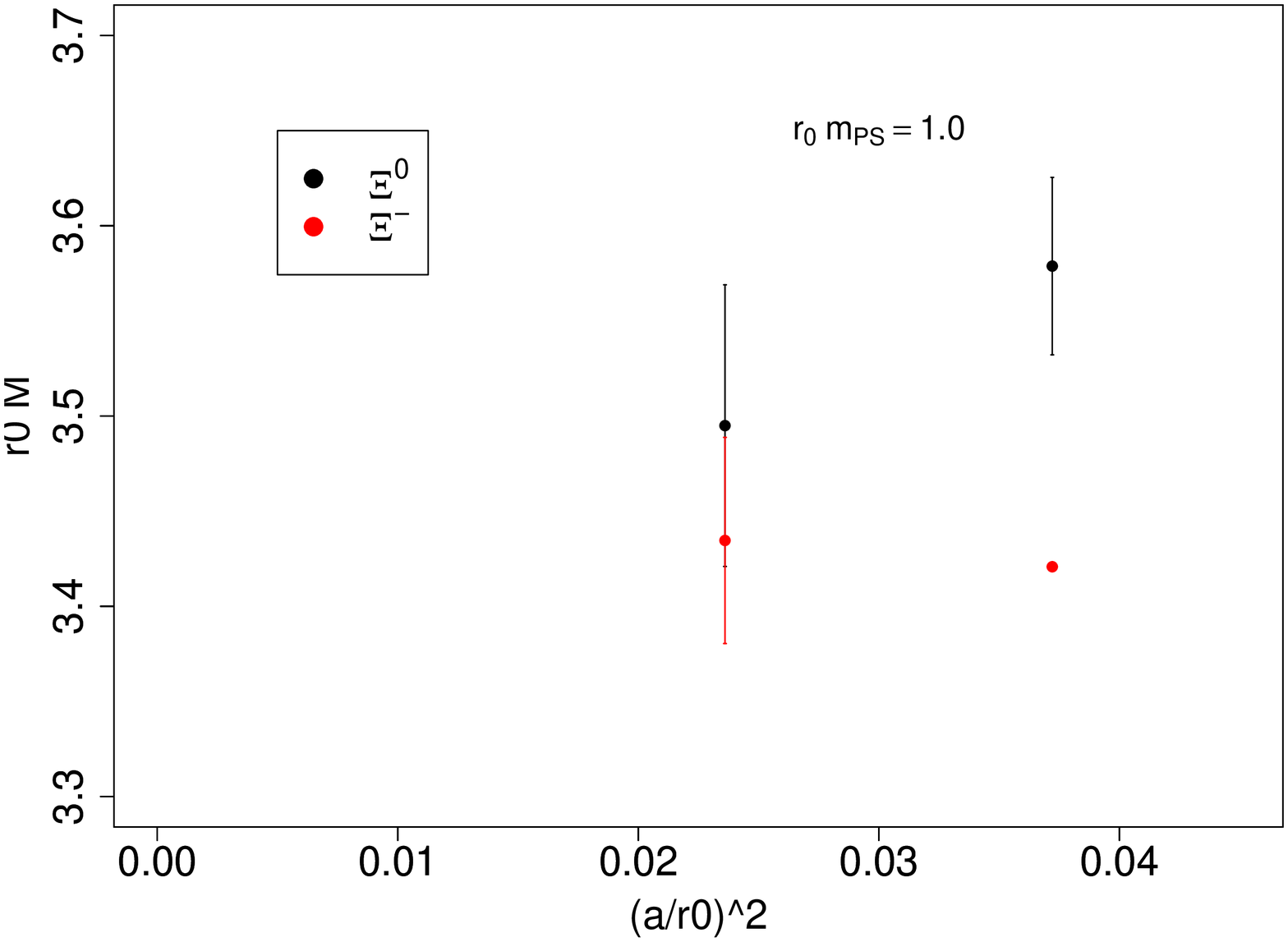}}}
\fi
\caption{$r_0 M_{\Xi}$ as a function of $(a/r_0)^2$ for $\Xi^0$ and  $\Xi^-$.}
\label{fig:artefact_xi}
\end{minipage}
\end{figure}
\end{center}

An important issue one needs to check in the twisted fermions formulation is the isospin 
symmetry breaking at finite lattice spacing.
 In the case of the $\Delta$ baryon it was shown in Ref.~\cite{nucleon} 
that the isospin breaking is compatible with zero.
In the strange baryon sector we can study the isospin splitting of 
the three  $\Sigma$ states and the two $\Xi$ states. 
In Figs.~\ref{fig:artefact_sigma} and \ref{fig:artefact_xi}
we show  the mass splitting for the $\Sigma's$ and $\Xi's$ 
respectively as a function of the lattice spacing.
 Using the value of the Sommer scale $r_0(a)$ at the chiral limit, 
we compare results obtained at different lattice spacings,
 for a fixed pion mass of reference ($r_0 m_{\pi} = 1.0$). 
We plot the behavior of $r_0 M_{\Sigma}$ and $r_0 M_{\Xi}$ 
as a function of  $(a/r_0)^2$. 
Results show a decrease of the isospin splitting with $a$.
 Furthermore, averaging over the different charge states of the
 $\Sigma$ and $\Xi$, results in a weaker lattice spacing dependence
 and justifies the use of the average for the chiral extrapolation.

\begin{figure}[ht]   
\begin{minipage}{7.4cm}
\ifpdf
{\mbox{\includegraphics[height=5.5cm,width=7.5cm]{./plots/Isospin_breaking_sigma}}}
\else
{\mbox{\includegraphics[height=5.5cm,width=7.5cm]{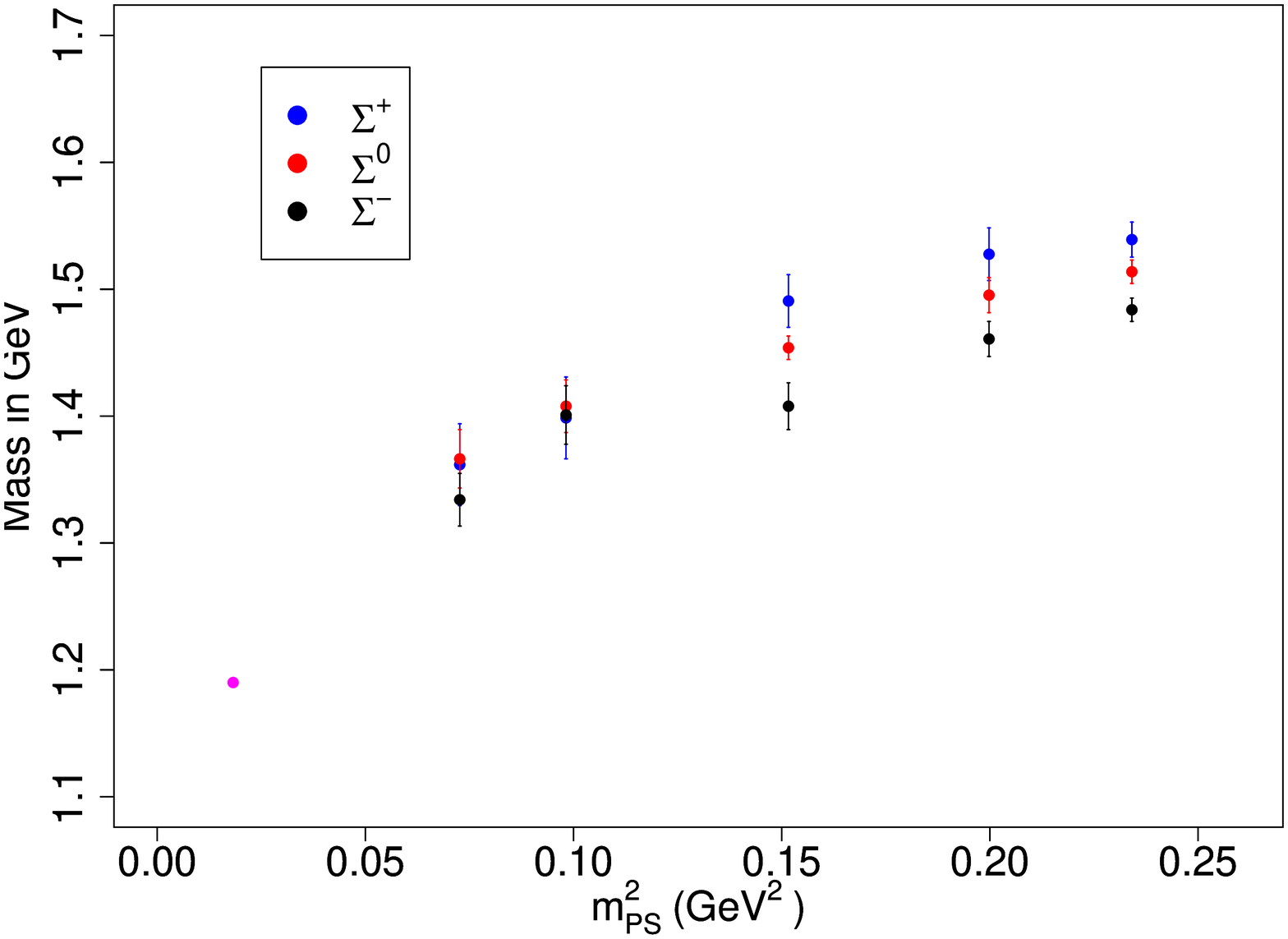}}}
\fi
\caption{ $\Sigma^{+}, \Sigma^0, \Sigma^-$ for $\beta=3.9$ ($a = 0.0085~\textit{fm}$)  for smeared-smeared correlators as a function of the pion mass squared ($\textit{GeV}^2$) .}
\label{fig:Isosigma}
\end{minipage}\hspace*{0.3cm}
\begin{minipage}{7.7cm}
\ifpdf
{\mbox{\includegraphics[height=5.5cm,width=7.5cm]{./plots/Isospin_breaking_xi}}}
\else
{\mbox{\includegraphics[height=5.5cm,width=7.5cm]{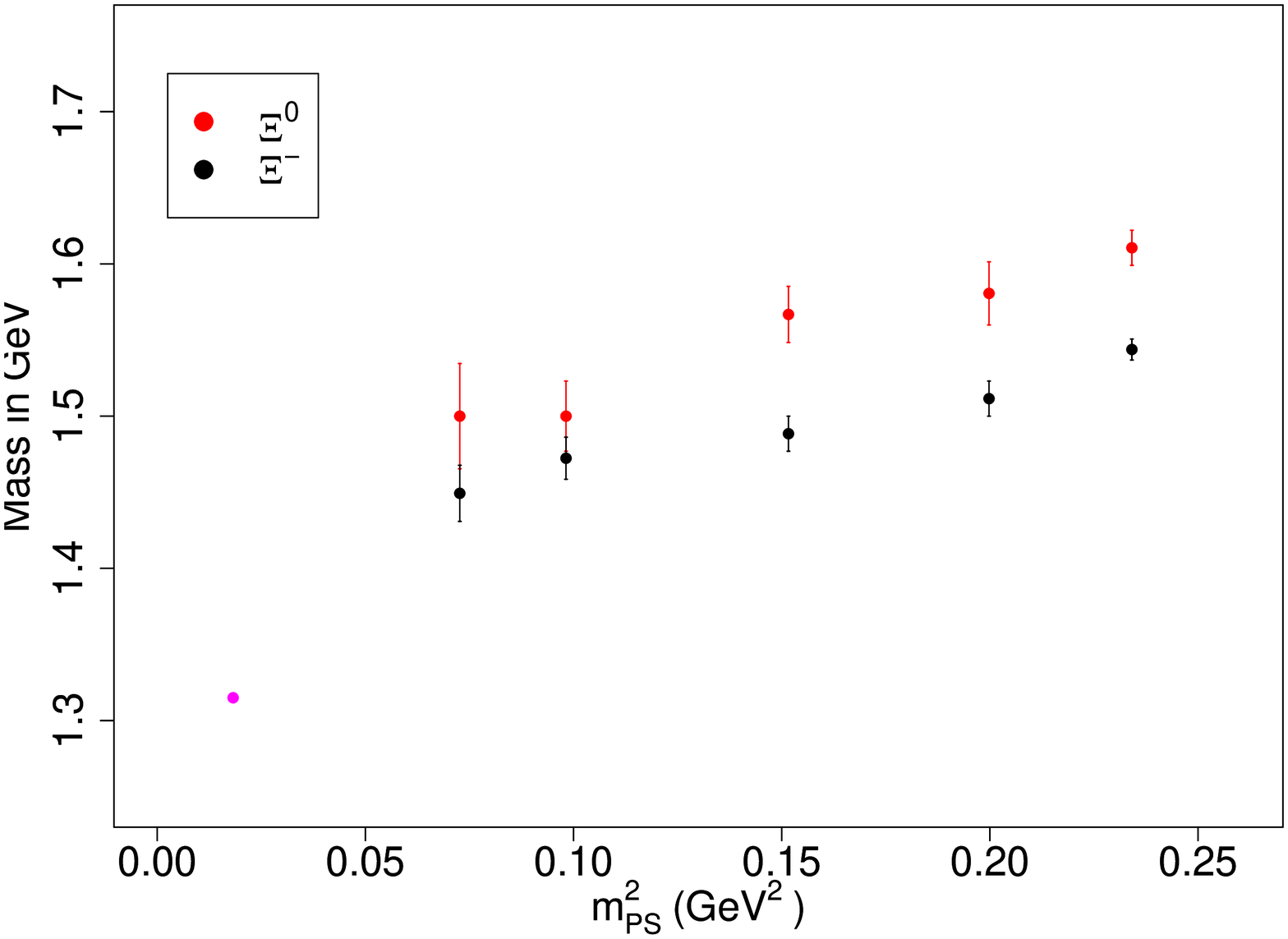}}}
\fi
\caption{$\Xi^0,\Xi^-$ for $\beta=3.9$ ($a = 0.0085~\textit{fm}$)  for 
smeared correlators as a function of the pion mass squared ($\textit{GeV}^2$) .}
\label{fig:Isoxi}
\end{minipage}
\end{figure}

In Figs.~\ref{fig:Isosigma} and \ref{fig:Isoxi} we display
 the mass of ($\Sigma^{+}, \Sigma^0, \Sigma^-$) and ($\Xi^0,\Xi^-$) multiplets 
 as a function of $m_{\pi}^2$ at fixed lattice spacing. 
We observe that the splitting decreases with the pion mass. 
The naive argument saying that for small quark mass, the u and d 
propagators are equivalent seems to apply in this case.

\section{$\Omega$ baryon}
\begin{figure}[ht]   
\begin{minipage}{7.5cm}
\ifpdf
{\mbox{\includegraphics[height=5.5cm,width=7.5cm]{./plots/s_dependance}}}
\else
{\mbox{\includegraphics[height=5.5cm,width=7.5cm]{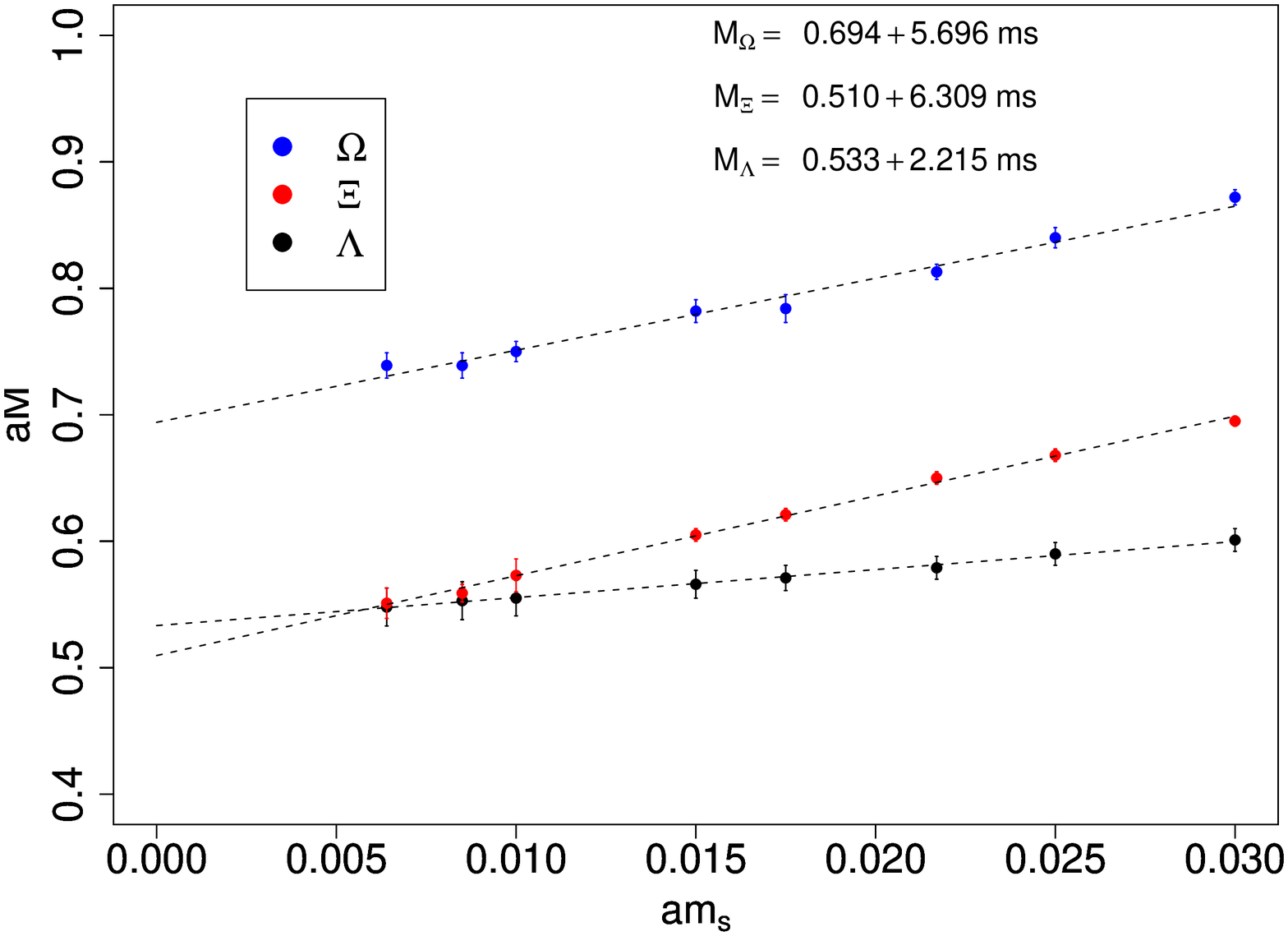}}}
\fi
\caption{Mass of $\Lambda$, $\Sigma$ and $\Omega$ at fixed lattice spacing for a pion mass of $\approx 310 \textit{MeV}$ as a function of the bare strange quark mass.}
\label{fig:s_dependance}
\end{minipage}\hspace*{0.3cm}
\begin{minipage}{7.5cm}
\ifpdf
{\mbox{\includegraphics[height=5.5cm,width=7.5cm]{./plots/nucleon_delta_omega}}}
\else
{\mbox{\includegraphics[height=5.5cm,width=7.5cm]{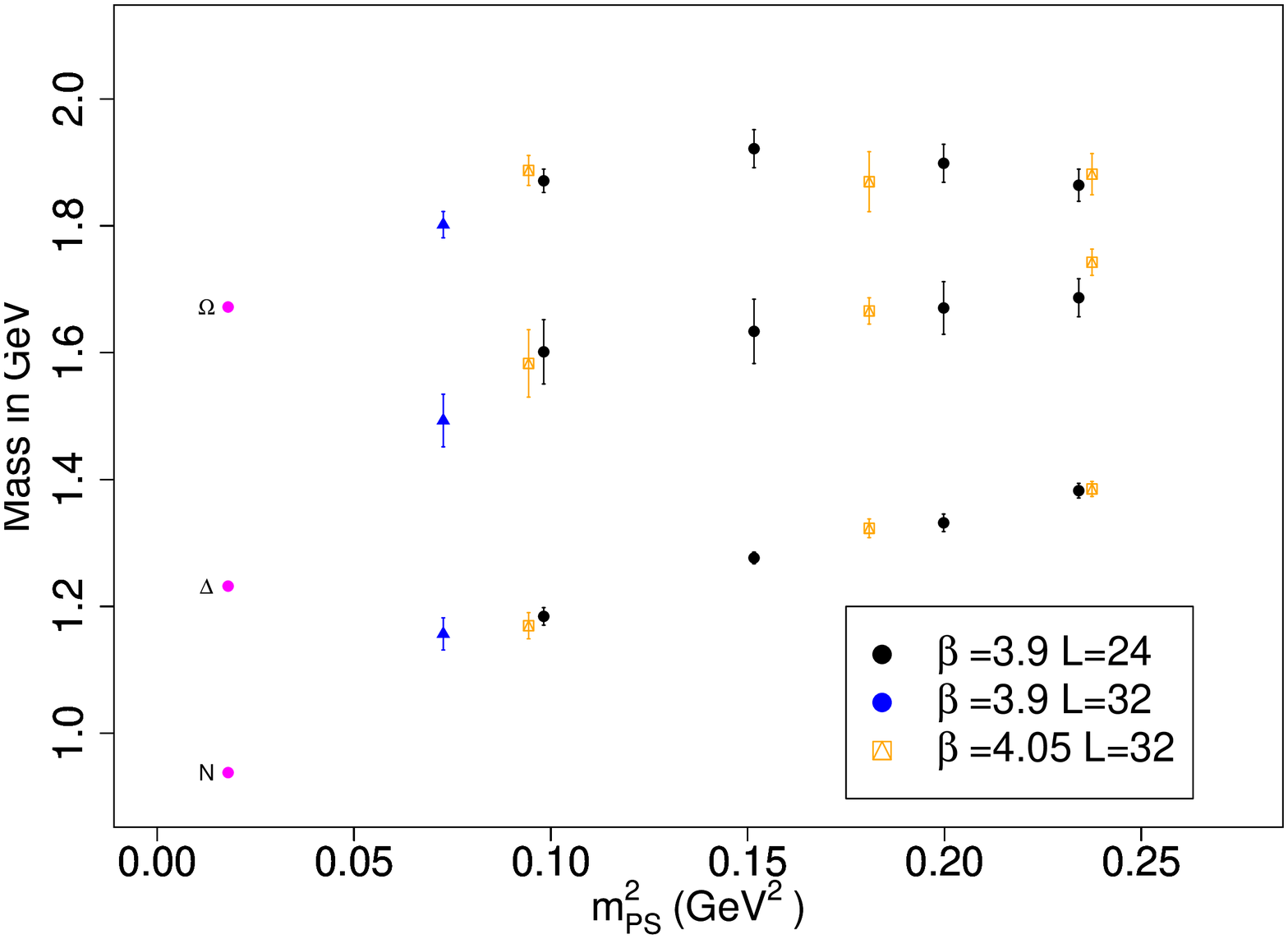}}}
\fi
\caption{Comparison of the pion mass dependence of $N$, $\Delta$ and $\Omega$, at fixed lattice spacing ($a=0.085\textit{fm}$). The experimental value
is shown magenta.}
\label{fig:nucleon_delta_omega}
\end{minipage}
\end{figure}

In this section we study the mass dependence of the $\Omega$, $\Xi$
and $\Lambda$ baryons on
the bare strange quark mass. In Fig.~\ref{fig:s_dependance},
 we show that their mass is, as expected,  linear as a function of $a\mu_s$.

The chiral extrapolation of the $\Omega$ mass
is very interesting because its dependence on  $m_{\pi}$ is a purely sea quark effect. 
Our data (see Fig.~\ref{fig:nucleon_delta_omega}) confirm that 
there is indeed a dependence on the u and d quark mass,
 which has to be understood in order to extrapolate the $\Omega$-mass 
to the physical point. 
As can be seen in Fig.~\ref{fig:nucleon_delta_omega}, the $\Omega$ 
cubic term is expected to be smaller than the nucleon and $\Delta$ one. 
A more extended discussion of the chiral extrapolation of the  $\Omega$-mass
will be detailed in a forthcoming publication~\cite{drach}.

\clearpage

\section{Conclusion} 
 We have shown, in this contribution, that the use of twisted mass fermions yields promising and accurate results in the spectroscopy of strange baryons. 
In the chiral extrapolation, the cubic term of the octet members can be constrained by studying their mass difference with the nucleon. This leads to the conclusion that the $m_{\pi}^3$ term in the octet chiral expansion is compatible with the nucleon one.
Contrary to the case of the $\Delta$, the strange baryon sector  shows an isospin breaking, which however decreases with the pion mass and becomes compatible with zero at $m_{\pi}\approx 310~\textit{MeV}$. A first study of lattice artifacts show small finite size effects. Strange quark mass dependence is, as expected, linear in $a\mu_s$. The $\Omega$ exhibits a sea quark dependence, which, even if it is small, shows the importance of dynamical quarks in this sector.

\section*{Acknowledgements}
The numerical calculations were performed at CC-IN2P3 and on computers of the CCRT (CEA) computing center. We thank them for their support.

\end{document}